\documentclass[12pt]{JHEP3}

\usepackage{amsmath,epsfig}
\usepackage{amssymb,amsfonts}
\usepackage{latexsym}
\usepackage[latin1]{inputenc}
\usepackage{subeqnarray}
\usepackage{xcolor}
\let\normalcolor\relax
\usepackage{graphicx}
\usepackage{longtable}
 
\relax
\renewcommand{\theequation}{\arabic{section}.\arabic{equation}}
\def\be{\begin{equation}}
\def\ee{\end{equation}}
\def\bea{\begin{eqnarray}}
\def\eea{\end{eqnarray}}

\newcommand\fverb{\setbox\pippobox=\hbox\bgroup\verb}
\newcommand\fverbdo{\egroup\medskip\noindent%
                        \fbox{\unhbox\pippobox}\ }
\newcommand\fverbit{\egroup\item[\fbox{\unhbox\pippobox}]}

\newcommand{\bear}{\begin{eqnarray}}

\newcommand{\eear}{\end{eqnarray}}

\newcommand{\bsea}{\begin{subeqnarray}}
\newcommand{\esea}{\end{subeqnarray}}
\newbox\pippobox

\def\6{\partial}

\def\a{\alpha}

\def\sp{\;\;\;,\;\;\;}

\def\p{\partial}

\def\sq
\def\a{\alpha}

\def\hri#1#2{\href{http://arxiv.org/abs/#1}{[ArXiv:#1]#2}}
\def\hre#1#2{\href{http://arxiv.org/abs/#1/#2}{[ArXiv:#1/#2]}}

\def\dd{{\rm d}}
\def\FF{{\cal F}}

\def\LL{{\cal L}}
\def\MM{{\cal M}}

\def\OO{{\cal O}}
\def\PP{{\cal P}}

\def\RR{{\cal R}}
\def\SS{{\cal S}}

\title{The holographic quantum effective potential at finite temperature and density}
\author{Elias Kiritsis$^{\flat,\natural}$, Vasilis Niarchos$^\flat$\\
~\\
$^\flat$ \href{http://hep.physics.uoc.gr}{Crete Center for Theoretical Physics},
Department of Physics,\\
University of Crete, 71003 Heraklion, Greece\\
~\\
$^\natural$ \href{http://www.apc.univ-paris7.fr}{APC, AstroParticule et Cosmologie}, Universit\'e Paris Diderot, CNRS/IN2P3, CEA/IRFU,
Observatoire de Paris, Sorbonne Paris Cit\'e,\\
 10, rue Alice Domon et L\'eonie Duquet, 75205 Paris
Cedex 13, France}

\preprint{CCTP-2012-12}

\abstract{We develop a formalism that allows the computation of the quantum effective potential of
a scalar order parameter in a class of holographic theories at finite temperature and charge density.
The effective potential is a valuable tool for studying the ground state of the theory, symmetry
breaking patterns and phase transitions. We derive general formulae for the effective potential and
apply them to determine the phase transition temperature and density in the scaling region.

}

\def\pdfannotlink{\pdfstartlink}
\begin{document}

\section{Introduction}
\label{intro}

Many recent applications of the AdS/CFT correspondence (in QCD or condensed matter theory)
entail the analysis of an effective holographic theory (EHT), namely a classical theory of gravity
coupled to a set of matter fields, $e.g.$ Einstein theory coupled to a set of real or complex scalar
fields and a set of gauge fields. Such an approach is the core of bottom-up approaches to
holographic problems. It was advocated in \cite{cgkkm} that this approach is well suited to study the
IR asymptotics of holographic theories and a host of new (generalized) classes of quantum criticality
were found at finite density. It is also in accord with alternative setups for the holographic flow in the
IR, \cite{holorg}.

The choice of interactions between the fields in the Lagrangian is limitless in bottom-up
approaches, and the usual strategy involves the choice of a specific well-motivated subset of
generic interactions parametrized by a number of arbitrary functions. For instance, the minimal set of
a single real or complex scalar field with a $U(1)$ gauge field has been the standard paradigm in
constructions of phenomenological models of YM \cite{ihqcd,gubser} and $s$-wave superconductors
at finite temperature and charge density, \cite{super}.

One of the central objects of interest in such systems is the free energy, which can be computed
using standard methods by evaluating the on-shell gravitational action. This quantity, as a function
of external sources, is related by a Legendre transform to the effective action of the theory. The
effective action contains all the information needed to determine the vacuum structure and other
generic correlation functions. An important part of the effective action is the (quantum) effective
potential. It is an important ingredient in deciding spontaneous symmetry breaking and superfluidity
(or superconductivity) at finite density.

Solving the full non-linear set of gravitational equations at finite temperature $T$ and charge density
$\rho$ (in generic systems of the type outlined above) is a highly non-trivial task that typically
requires numerical work. The purpose of this paper is to give compact expressions for
the quantum effective potential in holographic theories  in a form that facilitates further numerical
computation in concrete models. This is achieved by generalizing methods applied previously to zero
temperature and zero charge density solutions \cite{deboer,papa}. We introduce a properly defined
superpotential function $W$ and use it to reduce the second order differential equations to first order.
This reduction simplifies the differential equations that determine the background solution. The
quantum effective potential of the theory (which is determined from the on-shell gravitational action)
can be expressed compactly in terms of $W$ and the entropy function. We demonstrate how finite
temperature and charge density effects enter into this expression.

Perhaps the most appropriate way to think about the superpotential function is within the context of
the Hamilton-Jacobi treatment of holographic renormalization \cite{deboer,papa}. In this framework,
the on-shell gravitational action coincides with Hamilton's principal function $\SS_H$ which obeys a
first order differential equation, the so-called Hamilton-Jacobi equation. A solution of the
Hamilton-Jacobi equation determines a set of first order flow equations that specify the background
solution. In this context, the superpotential function $W$ is the potential term in $\SS_H$, and for
static and homogeneous solutions the first order equations are simply the Hamilton-Jacobi equation
for $W$ and the corresponding Hamilton-Jacobi flow equations for the fields.

This elegant reformulation of the second order equations of motion is generic and applies
also to finite-($T$, $\rho$) solutions. The practically difficult part of the implementation of this
formalism in this case is to determine the appropriate solution of the Hamilton-Jacobi equation.
Unfortunately, a generic ansatz for $\SS_H$ that captures finite temperature and density solutions,
even static and homogeneous ones, is not known and it is not obvious how to find it.
For that reason, in the present paper we propose a different less ambitious approach.

The approach that we adopt keeps the zero-($T$, $\rho$) first order flow equations in terms of a
function $W$ intact. With this ansatz we examine the extent to which the second order equations
reduce to a first order system and find that a full reduction to a first order system is possible for finite
$T$, but zero $\rho$. In the more general case of arbitrary $T$  and $\rho$ some equations
remain second order. The function $W$ that we define in this manner is the standard superpotential
that coincides with Hamilton's principal function $\SS_H$ in the $T=\rho=0$ case. For general
$T,\rho$, however, it does not coincide with $\SS_H$ and the relation between the $W$ that we
define and $\SS_H$ is less straightforward.

\subsection*{Summary of main results and structure of the paper}

We apply the above strategy to a rather general class of Einstein-Maxwell-dilaton theories
captured by the action \eqref{introaa}. These theories are relevant in holographic descriptions
of strange metals and other non-Fermi liquids and phenomenological models of the glue sector in
holographic QCD. The reduction of the second order equations of motion is discussed in section \ref
{1order} and the resulting equations are summarized in eqs.\ \eqref{introba}-\eqref{introbc}. A
compact general expression of the free energy at a general renormalization (RG) scale $M$ in 
terms of the function $W$ is derived in section \ref{effpotEMd} (eq.\ \eqref{effpotbe}). This equation is 
one of the main results of the paper.

The removal of  $M$ by taking $M\to\infty$ and the analysis of the corresponding
divergences is performed in sections \ref{zeromu}, \ref{finTmu} providing compact expressions for the
UV quantum effective potential. We use this result in section \ref{Tscale} to determine the critical
temperature $T_c$ at $\rho=0$ that separates a normal phase without a scalar condensate from an
ordered phase with a condensate. A general expression for the critical temperature at any charge
density is presented in section \ref{muscale}. It is known \cite{fhr} that this system can exhibit a
quantum critical point, $i.e.$ a critical point with $T_c=0$. We determine this point and related scaling
exponents analytically.

The case of Einstein-abelian Higgs actions with a complex scalar field, which are relevant
in discussions of $s$-wave holographic superconductors \cite{super} is more complicated.
 We have included the pertinent formulae in appendix \ref{EaH}.

As a preliminary illustration of a qualitatively different set of theories we consider analogous
computations in Einstein gravity coupled to a scalar field with a non-linear DBI-inspired action
(see eq.\ \eqref{introda}). Similar actions appear in discussions of the flavor sector in models
of holographic QCD \cite{tachyon}. A novelty in this case is the presence of multiple branches for the
superpotential function $W$. The relevant discussion, which focuses exclusively on
zero temperature, appears in appendix \ref{nonlinear}.

The detailed analysis of applications of this formalism is beyond the scope of the present
paper. A brief discussion of potential applications appears in the concluding section \ref{outlook}.
Additional useful technical details are relegated to the appendices \ref{UVexpansions}, \ref{expand}.

\section{Reduction of the equations of motion}
\label{1order}

In this section we describe the reduction of the second order gravitational equations
for a generic Einstein-Maxwell-dilaton model. We consider solutions at finite temperature and charge density.

\subsection{First order equations}
\label{1orderSub}

Our first example is provided by the following Einstein-Maxwell-dilaton action
in $d+1$ spacetime dimensions
\be
\label{introaa}
\begin{array}{rcl}
&&I=M_P^{d-1}\int_{\MM} \dd^{d+1}x \sqrt{-g} \left[ R- \frac{Z(\phi)}{4}F^2
-\frac{1}{2}(\partial \phi)^2 +V(\phi) \right]+I_{GH}
~,\\[3mm]
&&I_{GH}=2 M_P^{d-1}\int_{\partial \MM} \dd^dx \sqrt{h} K
~.
\end{array}
\ee
This action describes the dynamics of a $U(1)$ gauge field ${\bf A}_\mu$
(with field strength $F=d{\bf A}$) and a real scalar field $\phi$ coupled to Einstein gravity.
The boundary term $I_{GH}$ is the standard Gibbons-Hawking term needed to make
the variational problem well-defined. As such this action describes the grand canonical ensemble.
The Lagrangian is parameterized by two functions $(Z,V)$ of the scalar field $\phi$.
We assume that these functions have the following expansion around $\phi=0$
\be
\label{introab}
Z(\phi)=1+\sum_{n=1}^\infty z_n \phi^{2n}~, ~~
V(\phi)=\sum_{n=0}^{\infty} v_n \phi^{2n}
~.\ee

We are interested in solutions that asymptote to $AdS_{d+1}$. In the asymptotic region where
$\phi \to 0$ the first few coefficients of the expansion of the potential $V$ are
\be
\label{introac}
v_0=d(d-1)~, ~~ v_1=\frac{m^2}{2}
\ee
in units where the AdS radius is set to one.
$m$ denotes the mass of the scalar field $\phi$.
The standard AdS/CFT dictionary implies that the bulk scalar field $\phi$ maps
to a real scalar single-trace operator on the boundary with UV scaling dimension $\Delta$
such that
\be
\label{introad}
m^2=\Delta(d-\Delta)
~.
\ee
For $m^2$ in the range $(\frac{d^2}{4}-1,\frac{d^2}{4})$
there are two sensible values of $\Delta$ that obey this equation, $\Delta_\pm$
(by convention $\Delta_-=d-\Delta_+<\Delta_+$). In all other cases, only $\Delta_+$ is allowed.

Since we are interested in solutions with finite temperature and chemical potential we set
\be
\label{introaf}
\dd s^2=e^{2A(u)}\left( -f(u)\dd t^2+\dd x^i \dd x^i\right) +\frac{\dd u^2}{f(u)}~, ~~
{\bf A}=A_t(u)\dd t~, ~~
\phi=\phi(u)
~.\ee
We are working in domain wall frame coordinates where the AdS solution has
\be
\label{introag}
A(u)=-u ~, ~~
f(u)=1~, ~~ A_t(u)=0~, ~~ \phi(u)=0
~.
\ee
In general asymptotically AdS solutions the UV AdS boundary lies at $u\to -\infty$.

In this frame the second order equations of motion reduce to the following set of differential
equations
\begin{subequations}
\be
\label{introaia}
\frac{\dd}{\dd u}\left( e^{(d-2)A}Z \dot A_t\right)=0~,
\ee
\be
\label{introaib}
2(d-1)\ddot A+{\dot \phi}^2=0~,
\ee
\be
\label{introaic}
\ddot f+d \dot A \dot f-e^{-2A} Z {\dot A_t}^2=0~,
\ee
\be
\label{introaid}
(d-1)\dot A \dot f+\left(d(d-1) {\dot A}^2-\frac{1}{2} {\dot \phi}^2 \right)f-V
+\frac{1}{2} Ze^{-2A} {\dot A_t}^2=0
~.
\ee
\end{subequations}
We use the convention
\be
\label{introaj}
\dot ~=\frac{\dd}{\dd u}
~.
\ee

The first equation \eqref{introaia} can be solved trivially to obtain
\be
\label{introak}
\dot A_t=\frac{\rho}{e^{(d-2)A}Z}
~.
\ee
The integration constant $\rho$ is the charge density for the $U(1)$ gauge field.

By introducing a function $W$ such that
\be
\label{introal}
\dot \phi=W'(\phi)~, ~~ ':=\frac{\dd}{\dd \phi}
\ee
we notice that the second equation \eqref{introaib} is solved by setting
\be
\label{introam}
\dot A=-\frac{W(\phi)}{2(d-1)}
~.
\ee
Equivalently,
\be
\label{introan}
A(\phi)=A_0-\frac{1}{2(d-1)}\int_{\phi_0}^\phi \dd\tilde \phi\, \frac{W(\tilde\phi)}{W'(\tilde\phi)}~, ~~
u=u_0+\int_{\phi_0}^\phi \frac{\dd\tilde \phi}{W'(\tilde\phi)}
\ee
where $A_0=A(\phi_0)$.

$A_0$ is an arbitrary value of the scale factor at an intermediate position in the radial direction. Its 
choice corresponds to the definition of an RG scale $M=e^{A_0}$ at which the scalar operator is 
defined. Therefore $\phi_0$ is the effective ``source" field at the RG scale associated to $A_0$.

The third equation \eqref{introaic} takes the form
\be
\label{introao}
W'(W'f')'-\frac{d W W'}{2(d-1)}f'=\frac{\rho^2}{e^{2(d-1)A}Z}
~.\ee
This can be integrated once to obtain
\be
\label{introap}
W'f'=e^{-dA}\left[D+\rho^2 \int_{\phi_0}^\phi \frac{\dd\tilde \phi}{e^{(d-2)A}ZW'}\right]
~.\ee
$D$ is an integration constant that can be fixed in terms of the temperature $T$ and
the charge density $\rho$ by requiring regularity of the geometry at the horizon $u=u_h$
\be
\label{introaq}
\left| e^A \dot f \right|_{u=u_h}=4\pi T
~.
\ee
Choosing a negative sign for $\dot f$ at the horizon we obtain $\phi$ as a monotonically
increasing function of $u$.\footnote{In cases where we expand around an IR AdS solution
it is appropriate to pick a positive sign for $\dot f$.}
With this choice eq.\ \eqref{introaq} translates to
\be
\label{introar}
D+\rho^2\int_{\phi_0}^{\phi_h} \frac{\dd\tilde \phi}{e^{(d-2)A}ZW'}=-4\pi T \SS
\ee
where
\be
\label{introas}
\SS:=e^{(d-1)A(\phi_h)}=e^{(d-1)A_0-\frac{1}{2}\int_{\phi_0}^{\phi_h}\dd\tilde \phi \frac{W}{W'}}
=e^{(d-1)A_0}\, S~, ~~~ (\phi_h:=\phi(u_h))
\ee
is a quantity proportional to the entropy. Eq.\ \eqref{introap} becomes
\be
\label{introat}
W'f'=e^{-dA}\left[ -4\pi T\SS+\rho^2 \int_{\phi_h}^\phi \frac{\dd\tilde \phi}{e^{(d-2)A}ZW'} \right]
~.\ee
Integrating once (with the normalization condition\footnote{This condition imposes that the metric at the RG scale is the standard flat Minkowski metric.}   $f(\phi_0)=1$) we obtain
\be
\label{introau}
f(\phi)=1+\int_{\phi_0}^\phi \frac{\dd\tilde \phi}{W'}e^{-dA}
\left[ -4\pi T\SS+\rho^2 \int_{\phi_h}^\phi \frac{\dd\tilde \phi}{e^{(d-2)A}ZW'} \right]
~.\ee

Finally, the last equation \eqref{introaid} becomes
\be
\label{introav}
\left( \frac{d W^2}{2(d-1)}-{W'}^2\right) f-WW' f'=2V-\frac{\rho^2}{e^{2(d-1)A}Z}
~.
\ee

\subsubsection*{Summary of equations}

The ansatz \eqref{introal} has allowed us to reduce some of the second order equations to a
system of first order equations. A solution at finite temperature and charge density
can be found by solving (typically numerically) the system of coupled equations

\vspace{0.5cm}
\noindent
\begin{center}
\fbox{
\addtolength{\linewidth}{-30\fboxsep}%
\addtolength{\linewidth}{-30\fboxrule}%
\begin{minipage}{\linewidth}
\vspace{0.2cm}
\begin{subequations}
\be
\label{introba}
\RR' W'=\RR W~, ~~ \RR:=e^{-2(d-1)A}
\ee
\be
\label{introbb}
W'(W'f')'-\frac{d W W'}{2(d-1)}f'=\frac{\rho^2 \RR}{Z}
\ee
\be
\label{introbc}
\left( \frac{d W^2}{2(d-1)}-{W'}^2\right) f-WW' f'=2V-\frac{\rho^2\RR}{Z}
\ee
\end{subequations}
\vspace{0.1cm}
\end{minipage}
}
\end{center}
\vspace{0.5cm}

\noindent
for the unknown functions $\RR,f,W$.
The profile of the scalar field $\phi$ and gauge field
component $A_t$ can then be determined from eqs.\ \eqref{introak}, \eqref{introal}.

\subsection{Special cases and other comments}
\label{special}

In the special case of zero charge density, but arbitrary temperature, the second
equation \eqref{introbb} can be replaced by the first order differential equation
\eqref{introap}
\be
\label{introca}
W'f'=D e^{-dA}
~.
\ee
In this case we reduce the full set of equations to a set of first order equations.

Moreover, at zero temperature and charge density $f=1$ and the full solution can be determined
by solving a single first order equation for the function $W$
\be
\label{introcb}
\frac{dW^2}{2(d-1)}-{W'}^2=2V
~.
\ee
$W$ is the standard superpotential function in this case, the potential (derivative-independent) term
in Hamilton's principal function. In this language eq.\ \eqref{introcb} is the Hamilton-Jacobi
equation for static and homogeneous configurations.

In the more general case of non-zero temperature and charge density, the function $W$, as
defined here, does not express the full potential contribution to Hamilton's principal function. Solving
the Hamilton-Jacobi equation for static and homogeneous configurations at finite temperature
and charge density remains an interesting open problem.

Finally, a more general case of considerable interest for applications is the case
of the Einstein-abelian Higgs model expressed by the action
\be
\label{introcc}
I=M_P^{d-1}\int \dd^{d+1}x \sqrt{-g} \left[ R- \frac{Z(\phi)}{4}F^2
-\frac{1}{2}(\partial \phi)^2-
J(\phi) (\partial \theta-q {\bf A})^2+V(\phi) \right]+I_{GH}
~.\ee
This action describes the dynamics of a $U(1)$ gauge field ${\bf A}_\mu$
(with field strength $F=d{\bf A}$)
and a complex scalar field $\Phi:=\phi e^{i\theta}$ $(\theta\in [0,2\pi))$ coupled to Einstein gravity.
The constant $q$ denotes the $U(1)$ charge of the scalar field $\Phi$.

For solutions with a constant profile for the angular field $\theta$ we can consistently set
$\p_\mu \theta=0$ into the action to obtain
\be
\label{introcd}
I=M_P^{d-1}\int \dd^{d+1}x \sqrt{-g} \left[ R- \frac{Z(\phi)}{4}F^2-\frac{1}{2}(\partial \phi)^2-
q^2 J(\phi) {\bf A}^2+V(\phi) \right]+I_{GH}
\ee
which introduces a new set of interaction terms into the action \eqref{introaa}.
The equations of motion of this action and the first order reduction for solutions at finite
temperature and charge density in the approach of the previous subsection are summarized
in appendix \ref{EaH}.

\section{General expressions for the holographic effective potential}
\label{effpot}

\subsection{Introductory comments}

We are now in position to compute the effective potential. In order to fix the notation
let us recall first a few standard facts.

On the field theory side the free energy is defined, as a function of the sources, with a path integral
of the form
\be
\label{effpotaa}
Z[J]=e^{-\FF[J]}=\int e^{-\int d^dx \, \left( \LL + J\OO \right)}
\ee
where $\LL$ is the field theory Lagrangian. For concreteness, we focus here on a single scalar
operator $\OO$ and its source $J$. The vev of the operator $\OO$ is obtained from a
functional derivative of the free energy $\FF$ as follows
\be
\label{effpotab}
\langle \OO \rangle_J=\frac{\delta \FF}{\delta J}
~.
\ee

The effective action $\Gamma$, which is a function of the vev $\langle \OO \rangle$, is the Legendre
transform of the free energy $\FF$
\be
\label{effpotac}
\Gamma[\langle O \rangle_J]=\FF[J]-\int \dd^dx\, J(x) \langle \OO(x) \rangle_J
\ee
in terms of which the source can be expressed as
\be
\label{effpotad}
J=-\frac{\delta \Gamma}{\delta \langle \OO \rangle_J}
~.
\ee

On the gravity side, and the standard `quantization'\footnote{The so-called standard `quantization'
corresponds to the situation where the scalar operator $\OO$ has UV scaling dimension $\Delta_+$
in the notation below eq.\ \eqref{introad}. In the alternative `quantization' the operator $\OO$ has
UV scaling dimension $\Delta_-$.}
of the dual scalar field $\phi$, the free energy $\FF[J]$ is computed by evaluating the on-shell
gravitational action $I_{on-shell}$. The Legendre transform must then be implemented to obtain
the effective action $\Gamma$.
In the alternative `quantization', which is possible in the mass range
$m^2\in (\frac{d^2}{4}-1,\frac{d^2}{4})$, the on-shell gravitational action is automatically a function
of the vev $\langle \OO \rangle$ and expresses directly the effective action $\Gamma$.

For static and homogeneous configurations the effective action $\Gamma$ is proportional
to the effective potential $V_{eff}$
\be
\label{effpotae}
\Gamma[\langle \OO \rangle]=-\beta V_{d-1} V_{eff}(\langle \OO \rangle)
\ee
where $\beta$ is the period of the Wick rotated time direction and $V_{d-1}$ is the volume
of the $d-1$ spatial directions. The vacua of the theory are easily determined by extremizing the
effective potential, namely by solving the equations
\be
\label{effpotaf}
\frac{d\Gamma}{d\langle \OO\rangle}=0
~.
\ee
A vacuum is stable if and only if it is a minimum of the effective potential.

In the rest of this section we use the language of the previous section to
find general expressions of the on-shell gravitational action at finite temperature
and charge density.

\subsection{Free energy from gravity}
\label{effpotEMd}

The on-shell gravitational action $I$ has a bulk contribution $I_E$ and a boundary
contribution $I_{GH}$ from the Gibbons-Hawking term. Here we evaluate with a UV cutoff
at $u=u_0$. A standard computation gives the free energy
\be
\label{effpotbd}
\FF=I_{on-shell}=M_P^{d-1}\beta V_{d-1} \left(-W+\dot f \right)_{u=u_0}e^{d A_0}
~.\ee
Implementing the equations \eqref{introap} and \eqref{introar} we obtain the following expressions,
which will play a central role in this paper,
\be
\label{effpotbe}
\widehat \FF:=\frac{\FF}{M_P^{d-1}\beta V_{d-1}}=-e^{dA_0} W(\phi_0) +D
~,\ee
or equivalently,
\be
\label{effpotbf}
\widehat \FF=-e^{dA_0} W(\phi_0)
- 4\pi T\SS +\rho^2 \int_{\phi_h}^{\phi_0}\frac{\dd \tilde \phi}{e^{(d-2)A}ZW'}
~.\ee

By taking the RG scale $M=e^{A_0}\to\infty$ we recover the free energy as a function of the source (in the
case of the standard quantization) or directly the effective potential as a function of the
vev (in the case of the alternative quantization). This step and the resulting expressions
will be discussed in sections \ref{zeromu} and \ref{finTmu}.

\section{Renormalization group invariance}
\label{rg}

Holography provides a direct way of accessing the concept of RG running,
\cite{deboer}, and the analogue of $\beta$-functions, \cite{ihqcd}. In our case the $\beta$ function
for the scalar coupling $\phi$ can be obtained from (\ref{introal}), (\ref{introam}) to be
\be
\label{rgaa}
{\dd \phi\over \dd\log M}:= {\dd \phi \over \dd A}=\beta(\phi)\sp \beta(\phi)=-2(d-1)
\partial_{\phi}\log W(\phi)
\ee
where we identified $A$, in the standard manner, as the logarithm of the RG scale.

The effective potential for the sources, evaluated at $u=u_0$, is a function of the
RG scale $M=e^{A_0}$ and the sources $\phi_0$, but does not depend explicitly on $u_0$,
namely
\be
\label{rgab}
\frac{\p \widehat \FF}{\p u_0}=0
~.\ee
The RG running of the effective potential is controlled by the $\beta$-functions \eqref{rgaa}
\be
\label{rgac}
\frac{{\rm d} \widehat \FF}{{\rm d} A_0}=\frac{\p \widehat \FF}{\p A_0}
+\frac{\p \widehat \FF}{\p \phi_0} \beta(\phi_0)
\ee
which can be expressed further in terms of the superpotential function $W$.

For instance, in the zero-temperature case
\be
\label{rgae}
-{{\rm d}\widehat{\cal F} \over {\rm d}A_0}=
{{\rm d}\over {\rm d}A_0}\left[e^{dA_0} W(\phi_0)\right]=e^{dA_0}\left[dW+W'{d\phi_0\over dA_0}\right]=e^{dA_0}\left[dW-2(d-1){W'^2\over W}\right]
~.\ee
The superpotential equation (\ref{introav}) becomes at zero temperature and density
\be
\label{rgaf}
dW-2(d-1){W'^2\over W}=4(d-1){V\over W}
~.
\ee
Consequently,
\be
\label{rgag}
{\dd \widehat{\cal F} \over \dd A_0}=-4(d-1) e^{d A_0} \frac{V}{W}
~.
\ee

\section{Vanishing charge density}
\label{zeromu}

The case of zero temperature and density is well studied. Expressions for the effective action
in this case can be found for example in \cite{papa,Papadimitriou:2007sj,Faulkner:2010fh}.
By setting $T=\rho=0$ in the general equation \eqref{effpotbf} we find
\be
\label{zeromuaa}
\widehat \FF=-e^{dA_0}W
\ee
which expresses the free energy directly in terms of the `superpotential' function $W$,
\cite{deboer,Bianchi:2001de,papa}.
In this section, we proceed to generalize this result at finite temperature keeping the
charge density $\rho$ zero. At a finite cutoff $u_0$ the general relation \eqref{effpotbf} becomes
\be
\label{zeromuaaa}
\widehat \FF=-e^{dA_0}W(\phi_0)-4\pi T \SS(\phi_0)
~.
\ee
The temperature enters in this expression in two ways. First, it enters as a trivial overall factor in
front of the second term in the rhs. Second, it enters non-trivially through $\phi_h$
(the value of the scalar field on the horizon), which appears explicitly in the definition of
$\SS$ in eq.\ \eqref{introas}.

Furthermore, one can derive the following useful identities between $\phi_h$, $T$, $\alpha$,
and the constant $D$. By differentiating and combining the equations of motion (specifically,
the equations \eqref{zeromuae}, \eqref{zeromuaf} and \eqref{zeromuag}) one can
express the function $f$ as
\be
\label{zeromubf}
f=\frac{2V+WD e^{-dA}}{\frac{dW^2}{2(d-1)}-{W'}^2}
~.
\ee
Consequently, at the horizon we find that the following equation must hold
\be
\label{zeromubg}
2V_h+W_h D e^{-d A_h}=0
~,\ee
where we have defined
\be
\label{zeromubi}
V_h:=V(\phi_h)~, ~~ W_h:=W(\phi_h)~, ~~ A_h:=A(\phi_h)
~.
\ee
Now we can use eq.\ \eqref{zeromubg} together with the relation
\be
\label{zeromubj}
D=-4\pi T e^{(d-1)A_h}
\ee
to eliminate $e^{A_h}$
\be
\label{zeromubk}
e^{A_h}=2\pi T \frac{W_h}{V_h}
\ee
and express $D$ in the computationally more convenient form
\be
\label{zeromubl}
D=-2(2\pi T)^d \left( \frac{W_h}{V_h} \right)^{d-1}
~.
\ee
This allows us to recast eq.\ \eqref{effpotbe} (or \eqref{zeromuaaa}) into the form
\be
\label{zeromubla}
\widehat \FF=-e^{dA_0} W(\phi_0)-2(2\pi T)^d \left( \frac{W_h}{V_h} \right)^{d-1}
~.
\ee

The above equations express the free energy at a finite cutoff $u_0$ where $\phi=\phi_0$.
In order to remove the cutoff and make contact with the boundary QFT data, $J$ and
$\langle \OO \rangle$, we should scale $u_0\to -\infty$ and $A_0\to +\infty$.
At the asymptotic boundary, where $u\to -\infty$, the leading order terms in the expansion of the
scalar field $\phi$ are
\be
\label{zeromuab}
\phi=\alpha \, e^{u \Delta_-}+\ldots+ \zeta \, e^{u \Delta_+}+\ldots
~.
\ee
Then, in the case of the standard quantization, the holographic dictionary dictates the relation
\be
\label{zeromuac}
\alpha=J~, ~~ \zeta= \langle \OO \rangle
\ee
and, according to the general relations \eqref{effpotab}, the asymptotic coefficients
$\alpha$, $\zeta$ obey the equations
\be
\label{zeromuad}
\zeta=\frac{\dd \widehat \FF}{\dd \alpha}~~{\rm or~equivalently}~~
\alpha=-\frac{\dd \Gamma}{\dd \zeta}
~.
\ee
This relation together with a UV boundary condition, that provides an additional equation between
$\alpha$ and $\zeta$, determines the vacuum of the theory (and the full bulk solution) completely.

In the case of the alternative quantization the roles of leading and subleading coefficients
is exchanged. We proceed assuming the standard quantization.

\subsection{UV expansions}
\label{UVzeromu}

In the case at hand, a solution of the bulk equations is determined by solving the system of
first order equations\footnote{There is another class of potentials with
 exponential asymptotics related to decompactification limits in string theory, \cite{cgkkm}.
The superpotential for such asymptotics was analyzed in detail at finite
 temperature in \cite{gkmn} and at finite temperature and density in   \cite{cgkkm}.} 
\begin{subequations}
\be
\label{zeromuae}
\left( \frac{dW^2}{2(d-1)}-{W'}^2 \right)f-WW'f'=2V~,
\ee
\be
\label{zeromuaf}
W'f'=De^{-dA}
~,
\ee
\be
\label{zeromuag}
A=A_0-\frac{1}{2(d-1)}\int_{\phi_0}^\phi \dd \tilde \phi \frac{W}{W'}
~.
\ee
\end{subequations}

Useful information can be obtained by solving these equations perturbatively near $\phi=0$ in the
UV AdS asymptotic region. Given the expansion \eqref{introab} we set
\be
\label{zeromuaj}
W=\sum_{n=0}^\infty W_n(\phi)~, ~~ f=1+\sum_{n=1}^\infty f_n(\phi)~, ~~
e^{-pA}=e^{-pA_0}\sum_{n=1}^\infty g_n(\phi)
\ee
where $W_n,f_n,g_n$ have a separate expansion of the form
\be
\label{zeromuak}
W_n,f_n,g_n = \phi^{n\delta} \sum_{m=0}^\infty A_{n,m}\phi^m
~.\ee
Later we will fix $\delta=\frac{d}{\Delta_-}$.
We have assumed $\Delta_-<\frac{d}{2}$. In the special case where $\Delta_- = \Delta_+=\frac{d}{2}$
the latter expansion is performed in powers of $\log \phi$ instead of $\phi$. We will not consider
explicitly this case here (see, however, \cite{Amsel:2011km} for a related discussion).

Up to next-to-leading order we obtain
\begin{subequations}
\be
\label{zeromual}
\frac{d W_0^2}{2(d-1)}-{W_0'}^2=2V~,
\ee
\be
\label{zeromuam}
W_0'f_1'=De^{-d A_0} g_1~,
\ee
\be
\label{zeromuan}
\frac{d}{d-1}W_0 W_1-2 W_0' W_1'+2V f_1-W_0 W_0' f_1'=0
~.
\ee
\end{subequations}

The solution to the first equation, $W_0$, is the same as $W$ in the zero temperature case.
Soon we will see that $W_0$ is the term that controls the UV divergences. In that respect,
\eqref{zeromual} is in agreement with the fact that UV divergences are insensitive to the
temperature. It is known that there are two separate solutions to \eqref{zeromual} with
perturbative expansion
\be
\label{zeromuao}
W_0^{(\pm)}=2(d-1)+\frac{\Delta_\pm}{2}\phi^2
+\left( \frac{v_4}{d-4\Delta_\pm}-\frac{d \Delta_\pm^2}{16(d-1)(d-4\Delta_\pm)}\right)\phi^4+
\OO(\phi^6)
~.
\ee
It is also known \cite{papa,Papadimitriou:2007sj} that the $W_0^{(+)}$ solutions do not allow for
non-zero sources, whereas
the $W_0^{(-)}$ ones do. Hence, in what follows we will restrict our discussion to the $W_0^{(-)}$
solutions and will drop the superscript $^{(-)}$.

The next order correction $W_1$ takes the form
\be
\label{zeromuau}
W_1=W_1^{(T=0)}
+g_1 \int_{\phi_0}^\phi \dd\tilde \phi \left( \frac{V}{W_0'}\frac{f_1}{g_1}-D e^{-d A_0}
\frac{W_0}{2 W_0'} \right)
~.
\ee
The first term on the rhs of this equation, $W_1^{(T=0)}$, is the zero-temperature value of $W_1$
\be
\label{zeromuaz}
W_1^{(T=0)}=C \phi^{\frac{d}{\Delta_-}}
\left[ 1+
\left( \frac{d(d-2\Delta_-)}{4(d-1)(d-4\Delta_-)}-\frac{2d v_4}{\Delta_-^2(d-4\Delta_-)}\right)\phi^2
+\OO(\phi^4) \right]
\ee
and the second term a temperature-dependent contribution whose explicit form is determined
in appendix \ref{nochargeexpansions}. In \eqref{zeromuaz} $C$ is a constant fixed by infrared
regularity to a particular model-dependent value that we will denote as $C_*$.

\subsection{The UV region $A_0\to +\infty$ and renormalization}

As  $u_0\to -\infty$, $\phi_0$ behaves at leading order
as (see eq.\ \eqref{zeromuab})
\be
\label{zeromuba}
\phi_0=\alpha \, e^{u_0 \Delta_-}+\ldots
~.
\ee
At the same time, in order to have a regular limit for
\be
\label{zeromubb}
e^A=e^{A_0} \left( \frac{\phi_0}{\phi} \right)^{\frac{1}{\Delta_-}} \left( 1+\ldots \right)
=e^{A_0+u_0} \left( \frac{\alpha}{\phi} \right)^{\frac{1}{\Delta_-}} \left( 1+\ldots \right)
\ee
we require $A_0=-u_0\to +\infty$.

Then, one can check the following limits
\begin{subequations}
\be
\label{zeromubc}
\lim_{u_0\to -\infty}e^{dA_0} W_0=\infty
~,
\ee
\be
\label{zeromubd}
\lim_{u_0\to -\infty}e^{dA_0} W_1=C_* \alpha^{\frac{d}{\Delta_-}}-\frac{d-1}{d} D
~,
\ee
\be
\label{zeromube}
\lim_{u_0 \to -\infty} e^{dA_0} W_{n>1}=0
~.
\ee
\end{subequations}
The divergence in the free energy arises solely from the first term \eqref{zeromubc} that
can be removed by subtracting the zero-temperature superpotential, $W_C\big|_{T=0}$, for arbitrary
constant $C$. This subtraction removes the divergent piece and shifts the constant $C$ of
the $\alpha^{\frac{d}{\Delta_-}}$ term in $W_1$.
Therefore the renormalized superpotential is
\be
W_R(\phi,T)=W(C_*,T,\phi)-W_0(C,0,\phi)
\ee
Note that the renormalized superpotential $W_R$ depends on the arbitrary parameter $C$ used in the subtraction. This is a standard ``scheme dependence" as in renormalized QFT.

Using the expansion \eqref{zeromuap} we find that as we remove the
cutoff the function $S=e^{-\frac{1}{2}\int_{\phi_0}^{\phi_h} \dd \tilde \phi \frac{W}{W'}}$
takes the form
\be
\label{zeromubm}
S=\phi_0^{\frac{d-1}{\Delta_-}}\phi_h^{-\frac{d-1}{\Delta_-}} e^{\PP(\phi_h)}
\ee
where $\PP(\phi)$ is a function with an analytic expansion in powers of $\phi$.
Then,
\be
\label{zeromubn}
e^{-dA_h}=e^{-d A_0}S^{-\frac{d}{d-1}}\to \alpha^{-\frac{d}{\Delta_-}}\phi_h^{\frac{d}{\Delta_-}}
e^{-\frac{d}{d-1}\PP(\phi_h)}
\ee
and eq.\ \eqref{zeromubg} becomes
\be
\label{zeromubo}
2\pi T \alpha^{-\frac{1}{\Delta_-}}=\phi_h^{-\frac{1}{\Delta_-}} e^{\frac{1}{d-1}\PP(\phi_h)}
\frac{V_h}{W_h}
~.
\ee
The significance of this relation is the following. By inverting it we establish that
$\phi_h$ is a function of the dimensionally proper combination $2\pi T \alpha^{-\frac{1}{\Delta_-}}$.

We conclude that when we send $A_0\to +\infty$ (and subtract the divergence)
the free energy \eqref{zeromubla} at finite temperature (expressed in terms of bare UV variables)
takes the form
\be
\label{zeromubp}
\widehat \FF(\alpha)=
(C-C_*) \alpha^{\frac{d}{\Delta_-}}-\frac{2(2d-1)}{d}(2\pi T)^d \left( \frac{W_h}{V_h} \right)^{d-1}
~.
\ee
Hence, in the standard quantization the effective potential $V_{eff}(\zeta)$ can be determined
from the Legendre transform
\begin{subequations}
\be
\label{zeromubq}
V_{eff}(\zeta)=-(C-C_*) \alpha^{\frac{d}{\Delta_-}}
+\frac{2(2d-1)}{d}(2\pi T)^d \left( \frac{W_h}{V_h} \right)^{d-1}
+\alpha \zeta
~,
\ee
\be
\label{zeromubr}
\zeta=\frac{\dd \widehat \FF}{\dd \alpha}
~.
\ee
\end{subequations}
In the alternative quantization $\alpha$ represents the vev of the dual operator and
the rhs of eq.\ \eqref{zeromubp} expresses directly the effective potential. These expressions
are one of the main results of this paper.

Notice that the $T=0$ part of the free energy \eqref{zeromubp}, proportional to
$\alpha^{\frac{d}{\Delta_-}}$, follows also from the scale invariance of the planar boost-invariant
solution and agrees with the expressions in \cite{Faulkner:2010fh}. We will discuss the temperature
dependence further in section \ref{scaling}.

\section{Finite temperature at finite density}
\label{finTmu}

In this section we extend the discussion to arbitrary temperature and charge density.

We are now solving the more complicated system of first and second order differential equations
\eqref{introba}-\eqref{introbc}. The expansions \eqref{zeromuaj} have a slightly more involved form
which is presented in appendix \ref{chargeexpansions} along with the corresponding expansion of
the equations of motion.

One can show that the analog of eq.\ \eqref{zeromubf} is
\be
\label{finTmuba}
f=\frac{2V-\frac{\mu^2}{e^{2(d-1)A}Z}+e^{-dA} W\left( D+\rho^2 \int_{\phi_0}^\phi
\frac{1}{e^{(d-2)A}ZW'} \right)}
{\frac{d W^2}{2(d-1)}-{W'}^2}
~.
\ee
Hence, at the horizon $f(u_h)=0$ implies
\be
\label{finTmubb}
2V_h-\frac{\rho^2}{e^{2(d-1)A_h}Z_h}+e^{-dA_h} W_h \left( D+\rho^2 \int_{\phi_0}^{\phi_h}
\frac{1}{e^{(p-2)A}Z W'} \right)=0
~.
\ee
Implementing eq.\ \eqref{introar} we obtain
\be
\label{finTmubc}
2V_h-\frac{\rho^2}{e^{2(d-1)A_h}Z_h}-4\pi T {\cal S} e^{-d A_h} W_h=0
~.
\ee
Finally, with the use of eqs.\ \eqref{introas}, \eqref{zeromuba}, \eqref{zeromubm} we find
\be
\label{finTmubd}
2V_h-\frac{1}{Z_h} \rho^2 \alpha^{-\frac{2(d-1)}{\Delta_-}} \phi_h^{\frac{2(d-1)}{\Delta_-}}
e^{-2{\cal P}(\phi_h)}-4\pi T \alpha^{-\frac{1}{\Delta_-}} \phi_h^{\frac{1}{\Delta_-}}
e^{-\frac{{\cal P}(\phi_h)}{d-1}}W_h=0
~.
\ee
$\PP(\phi)$ is not identical to the function defined in eq.\ \eqref{zeromubm} for $\rho=0$, but is
defined in the same way.
Equation \eqref{finTmubd} is the generalization of eq.\ \eqref{zeromubo}. We conclude
that $\phi_h$ is now a function of the combinations
\be
\label{finTmube}
T\alpha^{-\frac{1}{\Delta_-}}~, ~~ \rho^2 \alpha^{-\frac{2(d-1)}{\Delta_-}}
~.
\ee

The free energy is given by eq.\ \eqref{effpotbf}
\be
\label{finTmuca}
\widehat \FF=-W(\phi_0)e^{pA_0}-4\pi T \SS
+\rho^2 \int_{\phi_h}^{\phi_0} \frac{1}{e^{(p-2)A}ZW'}
~.
\ee
As we remove the cutoff by taking $A_0\to +\infty$, and appropriately regulate the divergences
by subtraction as in the zero-charge case, we obtain
\be
\label{finTmucb}
-\lim_{A_0 \to \infty} W(\phi_0)e^{d A_0}
= (C-C_*) \alpha^{\frac{d}{\Delta_-}}+\frac{d-1}{d} D
~.
\ee
The sole contribution comes from $W_1$ (see eq.\ \eqref{finTmuaf}) as in the $\rho=0$ case. Recall,
however, that $D$ is now given in terms of the temperature by eq.\ \eqref{introar}. Hence, we
obtain
\be
\label{finTmucc}
\widehat \FF = (C-C_*) \alpha^{\frac{p}{\Delta_-}}
-\frac{2d-1}{d} 4\pi T \SS
-\frac{2d-1}{d} \rho^2 \int_{\phi_h}^{\phi_0} \frac{1}{e^{(d-2)A}ZW'}
~.
\ee

Using eq.\ \eqref{zeromubm} and defining for convenience the function
\be
\label{finTmucd}
\FF_\bullet (\phi_h):=\int_{\phi_h}^{\phi_0} \frac{1}{e^{(d-2)A}ZW'}
\ee
we find the expression
\be
\label{finTmuce}
\widehat \FF = (C-C_*) \alpha^{\frac{p}{\Delta_-}}
-\frac{2d-1}{d} 4\pi T \alpha^{\frac{d-1}{\Delta_-}} \phi_h^{-\frac{d-1}{\Delta_-}} e^{\PP(\phi_h)}
-\frac{2d-1}{d} \rho ^2 \FF_\bullet (\phi_h)
~.
\ee
The functions $\PP$, $\FF_\bullet$ are in general complicated model-dependent functions
that can be determined with the use of numerical methods.

\section{Scaling asymptotics and phase transitions}
\label{scaling}

The unwieldy model-dependent behavior of the general expression \eqref{finTmuce}
simplifies and reveals universal information in the vicinity of critical points. In this section
we discuss transitions that involve the `normal' phase with vanishing scalar condensate. For
simplicity, we consider the case of the alternative quantization in which $\alpha$ captures the
vev of the dual operator and the equations \eqref{zeromubp}, \eqref{finTmuce} express $V_{eff}$
directly as a function of $\alpha$.

\subsection{Vanishing charge density}
\label{Tscale}

As an interesting warmup exercise we first consider the case of vanishing charge density.

Solutions with everywhere small values of the scalar field $\phi$ are big black holes with
horizon in the UV region, namely $\phi_h\ll 1$. For such solutions the UV perturbative
expansions of subsection \ref{UVzeromu} are valid for the whole solution outside the horizon
and one can use them to evaluate perturbatively the rhs of eq.\ \eqref{zeromubp}.

First, inverting the relation \eqref{zeromubo} we find at leading order
\be
\label{Tscaleaa}
\phi_h=\left( \frac{4\pi T}{d} \right)^{-\Delta_-} \alpha+\ldots
~.
\ee
The dots indicate subleading terms in inverse powers of $T\alpha^{-\frac{1}{\Delta_-}}$.
From this expression we learn that in general $\phi_h \ll 1$ can be interpreted either as
the small vev limit at finite temperature or as the high temperature limit at fixed vev.

Inserting \eqref{Tscaleaa} into the expression for the effective potential and
expanding up to quadratic order we find
\be
\label{Tscaleab}
V_{eff}=-C\alpha^{\frac{d}{\Delta_-}}-(2d-1)\left( \frac{4\pi T}{d} \right)^d
- \frac{(2d-1)\Delta_-(d-2\Delta_-)}{4d}  \left( \frac{4\pi T}{d} \right)^{d-2\Delta_-} \alpha^2
+\ldots
\ee
where, compared to previous formulae, we have replaced the constant $C\to C+C_*$.
In the second, $\OO(\alpha^0)$, term we recognize the standard $T^d$ free energy of a
$d$-dimensional conformal field theory.

In the presence of a double-trace deformation on the field theory side
\be
\label{Tscaleac}
\delta \LL\sim g \OO^2
\ee
the effective potential at zero temperature becomes \cite{Faulkner:2010fh}
\be
\label{Tcab}
V_{eff}(\alpha)\big |_{T=0}=g \alpha^2-C\alpha^{\frac{d}{\Delta_-}}
~.
\ee
Hence, assuming for concreteness $C<0$, we learn that the normal vacuum at
$\alpha=0$ is unstable when $g<0$ and a stable symmetry-breaking vacuum exists
with vev
\be
\label{Tcaba}
\alpha=\left( \frac{2g \Delta_-}{d C}\right)^{\frac{\Delta_-}{d-2\Delta_-}}
~.
\ee

Adding temperature in the presence of the double-trace deformation we obtain the effective
potential
\be
\label{Tcac}
V_{eff}(\alpha)=-C\alpha^{\frac{d}{\Delta_-}}
-E T^d + g_{eff} \alpha^2+\ldots
\ee
where $g_{eff}$ is the temperature-shifted effective double-trace coupling
\be
\label{Tcad}
g_{eff}=g+G T^{d-2\Delta_-}
\ee
and $E$, $G$ are positive ($\alpha$, $T$)-independent constants that can be read off eq.\
\eqref{Tscaleab}.

The stability of the normal state at $\alpha=0$ can be determined immediately from the sign
of $g_{eff}$. In particular, the normal vacuum becomes unstable when $g_{eff}<0$. The critical
temperature $T_c$ that separates the stable from the unstable regime is obtained by setting
\be
\label{Tcae}
g_{eff}=0~~\Leftrightarrow ~~ T_c=\left(-\frac{g}{G} \right)^{\frac{1}{d-2\Delta_-}}
~.
\ee
This formula is in agreement with the result obtained in a different manner in \cite{fhr}.

\subsection{Finite temperature at finite density}
\label{muscale}

A similar analysis of the stability of the normal phase can be performed in the more general
case of finite temperature and density using the formulae of section \ref{finTmu}. Expanding
the effective potential up to quadratic order $\OO(\alpha^2)$ (in the presence of a double-trace
deformation) we obtain a formula analogous to \eqref{Tcac} with a more complicated effective
double-trace coupling. The explicit formula appears in eq.\ \eqref{muscaleaa} of appendix
\ref{expand}.

A finite-temperature transition at fixed $\rho$ can be determined again by setting $g_{eff}=0$.
This transition occurs at a quantum critical point when the equation $g_{eff}=0$ can be solved
for $T=0$. This is possible at the critical double-trace coupling
\be
\label{muscaleab}
g_c(\rho)=\frac{2d-1}{d}\rho^2
C_1 A_1^{\frac{d-2}{\Delta_-}}(\rho) \left(C_2 A_1^2(\rho) +\frac{d-2}{\Delta_-}A_2(\rho) \right)
~,
\ee
whose explicit $\rho$-dependence, and the constants $C_1,C_2$ are determined
in appendix \ref{expand}. To the best of our knowledge this analytic expression is new.
In the vicinity of the quantum critical point we observe the following scaling of the vev
\be
\label{muscaleac}
\langle \OO\rangle \sim \left( g_c-g \right)^{\frac{\Delta_-}{d-2\Delta_-}}
~.
\ee

\section{Outlook}
\label{outlook}

In quantum field theory the effective action contains all the information needed
to determine the vacuum structure and generic correlation functions of the theory. This information
can be used, for example, to identify critical points and phase transitions at finite temperature
and/or finite density, and as such it is very useful for descriptions of real world systems in
high energy physics or condensed matter. Unfortunately, the direct computation of the full effective
action of interacting QFTs is typically an almost impossible task. As a result, the available
effective theories are limited to descriptions of the vicinity of special points in parameter space.
For instance, once a critical point and an order parameter that characterizes the
corresponding transition is identified, a Landau-Ginzburg description can be employed to describe
the properties of this order parameter in the vicinity of the critical point.

Theories with a weakly curved dual gravitational description allow us to go beyond these limitations
and derive (with specific rules) the effective action of an order parameter in generic regimes of
parameters. In this work we have initiated a systematic study of the effective action
of theories with a dual description that involves a class of Einstein-Maxwell-dilaton theories.
We focused on spacetime-homogeneous configurations and the corresponding effective potential
at finite temperature and charge density. The resulting formulae extend the usual Landau-Ginzburg
treatments away from critical points and allow us to probe the full off-critical behavior of the system.
The latter is a rather complicated model-dependent function of the parameters. As expected, the
behavior simplifies near critical points where we can recover specific critical exponents and other
data of the transition.

There are several interesting extensions of this work that are worth exploring further.
First, it would be useful to explore the relation of our approach with the more systematic 
Hamilton-Jacobi
study of holographic renormalization \cite{deboer,papa}. For example, a more thorough
understanding of the precise relation between the superpotential function $W$ that we introduced
and Hamilton's principal function at finite temperature and density would be useful.

Second, for many applications it is useful to know the full effective action including derivative
interactions, a subject interesting for applications both to finite density systems and cosmology.
 Work in this direction using the Hamilton-Jacobi formalism has appeared in
\cite{Papadimitriou:2007sj,papa}, where explicit expressions at zero-temperature and density are
provided. In this work we focused on effective potentials for scalar order parameters, but more
generally effective actions for generic tensor order parameters ($e.g.$ vectors) are of intrinsic
interest in various applications.

The Einstein-Maxwell-dilaton theories in this paper have appeared in numerous applications
of the holographic duality to QCD and condensed matter theory. A class of these applications
in condensed matter theory refers to holographic superconductors \cite{fhr} and holographic
models of magnetism \cite{Iqbal:2010eh}. The effective potential that we analyzed above can be used
to study the vacuum structure of these, and more general, systems. These systems can also be used
as basic building blocks in the construction of corresponding Josephson junction networks within the
framework of designer multi-gravity theories \cite{Kiritsis:2011zq}. Applying the results obtained
above to this context we can determine the effective potential of corresponding network theories and
use it to study their finite temperature and density physics.

\section*{Acknowledgements}\label{ACKNOWL}
\addcontentsline{toc}{section}{Acknowledgements}

We thank Ioannis Papadimitriou for useful discussions.
This work was in part supported by the European grants
FP7-REGPOT-2008-1: CreteHEPCosmo-228644, PERG07-GA-2010-268246, and
the EU program ``Thalis'' ESF/NSRF 2007-2013.

\appendix
\renewcommand{\theequation}{\thesection.\arabic{equation}}
\addcontentsline{toc}{section}{Appendices\label{app}}
\section*{Appendices}

\section{The Einstein-abelian-Higgs model}
\label{EaH}

As explained in section \ref{special} the action of the Einstein-abelian-Higgs model
\eqref{introcc} reduces to the action \eqref{introcd}
\be
\label{EaHaa}
I=M_P^{d-1}\int \dd^{d+1}x \sqrt{-g} \left[ R- \frac{Z(\phi)}{4}F^2-\frac{1}{2}(\partial \phi)^2-
q^2 J(\phi) {\bf A}^2+V(\phi) \right]+I_{GH}
\ee
by setting $\partial_\mu \theta=0$. $\phi$ is a real scalar field that represents the modulus
of the complex scalar field with $U(1)$ charge $q$ of the Einstein-abelian-Higgs model.
In the present appendix we summarize the equations of motion of the action
\ref{EaHaa} and the form they take when a superpotential function $W$ is introduced as in section
\ref{1orderSub}.

In the domain wall frame
\be
\label{EaHab}
\dd s^2=e^{2A(u)}\left(-f(u) \dd t^2+\dd x^i \dd x_i\right)+\frac{\dd u^2}{f(u)}
~, ~~ {\bf A}=A_t(u)\dd t~, ~~ \phi=\phi(u)
\ee
we obtain the following independent set of second order differential equations
\begin{subequations}
\be
\label{EaHac}
\frac{\dd}{\dd u}\left(e^{(d-2)A}Z\dot A_t\right)-2q^2e^{(d-2)A}\frac{J}{f}A_t=0
~,
\ee
\be
\label{EaHad}
2(d-1)\ddot A+{\dot \phi}^2+2q^2e^{-2A}\frac{J}{f^2}A_t^2=0
~,
\ee
\be
\label{EaHae}
\ddot f+d \dot A \dot f-e^{-2A}Z{\dot A_t}^2-2q^2 e^{-2A}\frac{J}{f} A_t^2=0
~,
\ee
\be
\label{EaHaf}
(d-1)\dot A \dot f+\left( d(d-1) \dot A^2-\frac{1}{2}\dot \phi^2 \right) f-V
+\frac{1}{2}Z e^{-2A} {\dot A_t}^2 -q^2J e^{-2A}f^{-1} A_t^2=0
~.
\ee
\end{subequations}

We introduce the function $W$ by requiring the flow equation
\be
\label{EaHag}
\dot \phi=W'
~.
\ee
Then, eq.\ \eqref{EaHad} gives
\be
\label{EaHai}
\dot A=-\frac{1}{2(d-1)}\left( W+2q^2 \int_{\phi_0}^\phi \dd\tilde \phi \, e^{-2A}\frac{JA_t^2}{f^2 W'}
\right)
\ee
where we have chosen a particular integration constant that fixes the definition of $W$.
Eq.\ \eqref{EaHae} becomes
\be
\label{EaHaj}
W' (W' f')'-\frac{d}{2(d-1)}\left( W+2q^2\int^\phi_{\phi_0} \frac{e^{-2A} J}{f^2W'} A_t^2 \right)
W'f'-e^{-2A}Z{\dot A_t}^2-2q^2e^{-2A}\frac{J}{f}A_t^2=0
\ee
and eq.\ \eqref{EaHaf}
\bea
\label{EaHak}
&&-\left( W+2q^2 \int_{\phi_0}^\phi \dd\tilde \phi \, e^{-2A}\frac{JA_t^2}{f^2 W'}
\right)W'f'
\nonumber\\
&&+\left( \frac{d}{2(d-1)} \left( W+2q^2 \int_{\phi_0}^\phi \dd\tilde \phi \, e^{-2A}\frac{JA_t^2}{f^2 W'}
\right)^2- {W'}^2 \right)f
-2V
\nonumber\\
&&+Ze^{-2A}{\dot A_t}^2
-2q^2 Je^{-2A}f^{-1}A_t^2=0
~.
\eea

Eq.\ \eqref{EaHaj} can be integrated to obtain the expression
\be
\label{EaHal}
W'f'=e^{-dA}\left( D+e^{dA}H\int_{\phi_0}^\phi \dd\tilde \phi\,
\frac{e^{-2A}\left( Z{\dot A_t}^2+\frac{2q^2J}{f} A_t^2 \right)}{H W'} \right)
~,
\ee
where $H(\phi)$ is the function that solves the differential equation
\be
\label{EaHam}
H'=\frac{d}{2(d-1)}\frac{W+2q^2\int_{\phi_0}^\phi \dd\tilde \phi e^{-2A}\frac{J}{f^2 W'}A_t^2}{W'} H
\ee
and $D$ is an integration constant. Using the horizon regularity condition
\be
\label{EaHan}
\left | e^A \dot f \right |_{u=u_h}=4\pi T
\ee
we determine $D$ as follows
\be
\label{EaHao}
D+e^{dA(u_h)}H(u_h)\int_{\phi_0}^{\phi_h}
\frac{e^{-2A}(Z{\dot A_t}^2+\frac{2q^2J}{f}A_t^2)}{HW'}=-4\pi T e^{(p-1)A(u_h)}=-4\pi T \SS
~.
\ee

The free energy takes the `universal' form \eqref{effpotbe}
\be
\label{EaHap}
\frac{\FF}{M_P^{d-1}\beta V_{d-1}}=-e^{dA_0} W(\phi_0) +D
~.
\ee

\section{UV expansions}
\label{UVexpansions}

\subsection{UV expansions for vanishing charge density}
\label{nochargeexpansions}

In this appendix we consider in more detail the solution of the equations \eqref{zeromuam},
\eqref{zeromuan}
\begin{subequations}
\be
\label{zeromuamApp}
W_0'f_1'=De^{-d A_0} g_1~,
\ee
\be
\label{zeromuanApp}
\frac{d}{d-1}W_0 W_1-2 W_0' W_1'+2V f_1-W_0 W_0' f_1'=0
~.
\ee
\end{subequations}

From eq.\ \eqref{zeromuag} and the expansion
\bea
\label{zeromuap}
&&\frac{d}{2(d-1)}\int_{\phi_0}^\phi \dd \tilde \phi \frac{W}{W'}=
\nonumber\\
&&=\frac{d}{2(d-1)}\int_{\phi_0}^\phi \dd \tilde \phi \frac{W_0}{W'_0}
\left[ 1+\left(\frac{W_1}{W_0}-\frac{W'_1}{W_0'}\right)
+\left( \frac{W_2}{W_0}-\frac{W_2'}{W_0'}+\frac{{W_1'}^2}{{W_0'}^2} \right)+\ldots\right]=
\nonumber\\
&&=\log g_1+\frac{g_2}{g_1}+\frac{g_3}{g_1}-\frac{g_2^2}{2g_1^2}+\ldots
\eea
we obtain
\begin{subequations}
\be
\label{zeromuaq}
g_1=e^{\frac{d}{2(d-1)}\int_{\phi_0}^\phi \dd \tilde \phi \frac{W_0}{W_0'}}
~,
\ee
\be
\label{zeromuar}
g_2=\frac{d}{2(d-1)}g_1\int_{\phi_0}^\phi \dd\tilde \phi \frac{W_0}{W_0'}
\left( \frac{W_1}{W_0}-\frac{W_1'}{W_0'}\right)
~, ~~{etc.}
\ee
\end{subequations}
$W_0$ is a solution of eq.\ \eqref{zeromual} with the $(-)$ expansion in \eqref{zeromuao}.

We can integrate eq.\ \eqref{zeromuam} to find $f_1$
\be
\label{zeromuas}
f_1=De^{-d A_0}\int_{\phi_0}^\phi \dd\tilde \phi\frac{g_1}{W_0'}
~.
\ee
Then eq.\ \eqref{zeromuan} becomes
\be
\label{zeromuat}
W_1'-\frac{d}{2(d-1)}\frac{W_0}{W_0'}W_1=\frac{V}{W_0'}f_1-\frac{W_0}{2}f_1'
\ee
with solution
\be
\label{zeromuauApp}
W_1=W_1^{(T=0)}
+g_1 \int_{\phi_0}^\phi \dd\tilde \phi \left( \frac{V}{W_0'}\frac{f_1}{g_1}-D e^{-d A_0}
\frac{W_0}{2 W_0'} \right)
\ee
where $W_1^{(T=0)}$ is the zero-temperature value of $W_1$
\be
\label{zeromuazApp}
W_1^{(T=0)}=C \phi^{\frac{d}{\Delta_-}}
\left[ 1+
\left( \frac{d(d-2\Delta_-)}{4(d-1)(d-4\Delta_-)}-\frac{2d v_4}{\Delta_-^2(d-4\Delta_-)}\right)\phi^2
+\OO(\phi^4) \right]
~.
\ee
As was commented in the main text, $C$ is a constant fixed by infrared regularity to a particular
model-dependent value that we denote as $C_*$.

\subsection{UV expansions at finite charge density}
\label{chargeexpansions}

At finite charge density we are solving the more complicated system of first and second order
differential equations \eqref{introba}-\eqref{introbc}.
The expansions \eqref{zeromuaj} are slightly more involved
\be
\label{finTmuae}
h=\sum_{n,m} h_{n,m}(\phi) \phi^{\frac{dn+2m(d-1)}{\Delta_-}}~, ~~
h_{n,m}(\phi)=\sum_{\ell=0}^\infty h_{n,m,\ell}\phi^\ell~, ~~
{\rm for}~ h=(W,f,\RR)
~.
\ee
For the first few orders
\begin{subequations}
\be
\label{finTmuaf}
W=W_0+W_1+\widetilde W_1+\ldots~, ~~
W_1= \phi^{\frac{d}{\Delta_-}}\sum_{n=0}^\infty W_{1,n}\phi^n ~, ~~
\widetilde W_1= \phi^{\frac{2(d-1)}{\Delta_-}} \sum_{n=0}^\infty \widetilde W_{1,n} \phi^n
~,
\ee
\be
\label{finTmuag}
\RR=\RR_0+\RR_1+\widetilde \RR_1+\ldots~, ~~
\RR_1= \phi^{\frac{d}{\Delta_-}}\sum_{n=0}^\infty \RR_{1,n}\phi^n ~, ~~
\widetilde \RR_1= \phi^{\frac{2(d-1)}{\Delta_-}} \sum_{n=0}^\infty \widetilde \RR_{1,n} \phi^n
~,
\ee
\be
\label{finTmuai}
f=f_0+f_1+\widetilde f_1+\ldots~, ~~
f_1= \phi^{\frac{d}{\Delta_-}}\sum_{n=0}^\infty f_{1,n}\phi^n ~, ~~
\widetilde f_1= \phi^{\frac{2(d-1)}{\Delta_-}} \sum_{n=0}^\infty \widetilde f_{1,n} \phi^n
~.
\ee
\end{subequations}
The form of the equations and the corresponding solution at $\rho=0$ suggests setting
\be
\label{finTmuaj}
\RR_0=\RR_1=0~, ~~ f_0=1
~.
\ee

Inserting these expansions into the equations of motion \eqref{introba}-\eqref{introbc}
we find that the functions $W_0, W_1, f_1$ are the same as in the $\rho=0$ case and
\begin{subequations}
\be
\label{finTmuak}
\widetilde \RR_1=e^{-2(d-1)A_0 +\int_{\phi_0}^\phi \frac{W_0}{W_0'}}
~,
\ee
\be
\label{finTmual}
\widetilde f_1'=\frac{\mu^2 g_1}{W_0'} \int_{\phi_0}^\phi \frac{\widetilde \RR_1}{g_1ZW_0' }
~,
\ee
\be
\label{finTmuam}
\widetilde W_1=g_1 \int_{\phi_{*}}^\phi \frac{1}{g_1}
\left( \frac{V}{W_0'}\widetilde f_1 -\frac{W_0}{2} \widetilde f_1'
+\frac{\mu^2 \widetilde \RR_1}{2 Z W_0'}
\right)
~,
\ee
\end{subequations}
where $g_1$ is the function that appears in eq.\ \eqref{zeromuaj}
and $\phi_*$ is an integration constant.

\section{Small-$\phi_h$ expansion details}
\label{expand}

In the regime $\phi_h \ll 1$ we have the following expansion
\be
\label{expandaa}
\phi_h =A_1 \alpha(1+A_2 \alpha^2+\ldots)~, ~~
\ee
\be
\label{expandab}
\PP(\phi_h)=B_1 \phi_h^2+\ldots
~,
\ee
\be
\label{expandac}
\FF_\bullet(\phi_h)=C_1\alpha^{-\frac{p-2}{\Delta_-}}\phi_h^{\frac{p-2}{\Delta_-}}
\left( 1+C_2 \phi_h^2 +\ldots \right)
\ee

$B_1$ can be determined from the definition \eqref{zeromubm}
\be
\label{expandad}
B_1=-\frac{d-16\Delta_-}{32(d-4\Delta_-)}
~.
\ee

The coefficients $A_1,A_2$ are determined by expanding eq.\ \eqref{finTmubd} in $\phi_h$.
More specifically, $A_1$ is determined by solving the algebraic equation
\be
\label{expandae}
\rho^2 A_1^{\frac{2(d-1)}{\Delta_-}}+8\pi(d-1) T  A_1^{\frac{1}{\Delta_-}}-2d(d-1)=0
~.
\ee
$A_2$ is determined in terms of $A_1$ as follows
\bea
\label{expandaf}
&&A_2=\frac{\Delta_-}{2(d-1)} \times
\\
&&\times
\frac{(2d(d-1)z_1+m^2 )A_1^2
+2B_1 \rho^2 A_1^{\frac{2(d-1)}{\Delta_-}+2}
-2\pi T A_1^{\frac{1}{\Delta_-}+2}(4(d-1)z_1+(\Delta_- -4B_1))}
{\rho^2 A_1^{\frac{2(d-1)}{\Delta_-}}+4\pi T A_1^{\frac{1}{\Delta_-}}}
~.\nonumber
\eea

Finally, the coefficients $C_1, C_2$ are determined by expanding the expression
\eqref{finTmucd} and taking the $\phi_0\to 0$ limit. We find
\be
\label{expandag}
C_1=-\frac{1}{d-2}
~,
\ee
\be
\label{expandai}
C_2=\frac{1}{16}\frac{d-16\Delta_-}{d-4\Delta_-}-z_1
-\frac{3}{\Delta_-(d-4\Delta_-)} \left( v_4-\frac{d\Delta_-^2}{16(d-1)} \right)
~.
\ee

Expanding the effective potential up to quadratic order $\OO(\alpha^2)$ (in the presence of a
double-trace deformation) we obtain a formula analogous to \eqref{Tcac} with a more complicated
effective double-trace coupling
\bea
\label{muscaleaa}
&&g_{eff}=g-
\frac{2d-1}{d} \Bigg[ (4\pi T)^d
\left( \frac{2 (d-1)}{2d(d-1)-\rho^2 A_1^{\frac{2(d-1)}{\Delta_-}}} \right)^{d-1}
\nonumber\\
&&\left( \frac{\Delta_- A_1^2}{4}
-(d-1)\frac{m^2 A_1^2-\rho^2 A_1^{\frac{2(d-1)}{\Delta_-}}
\left( \frac{2(d-1)A_2}{\Delta_-}-z_1A_1^2-2B_1 A_1^2\right)}
{2d(d-1)-\rho^2A_1^{\frac{2(d-1)}{\Delta_-}}}\right)
\nonumber\\
&&+\rho^2 C_1 A_1^{\frac{d-2}{\Delta_-}}\left(C_2 A_1^2+\frac{d-2}{\Delta_-}A_2\right)
\Bigg]
~.
\eea

\section{Non-linear Einstein-scalar actions}
\label{nonlinear}

As a qualitatively different example, in this appendix we consider actions of the
general form
\be
\label{introda}
I=M_P^{d-1}\int \dd^{d+1}x\, \sqrt{-g}\left[ R+V(\phi) F\left( (\p \phi)^2 \right) \right]+I_{GH}
~.
\ee
$\phi$ is a real scalar field. For the function $F(x)$ we demand $F(0)=1$, $F'(0)=-\frac{\xi^2}{2}<0$
so that in the long-wavelength limit we obtain the canonical scalar field action
\bea
\label{introdb}
&&I=M_P^{d-1}\int \dd^{d+1}x\, \sqrt{-g}\left[ R+\widehat V(\chi)-\frac{1}{2}(\p \chi)^2+\ldots \right]
+I_{GH}~
\nonumber\\
&&\chi=\xi\int_0^\phi \dd x \sqrt{V(x)}~, ~~ \widehat V(\chi):=V(\phi(\chi))
~.
\eea
A commonly encountered example is the DBI case with
\be
\label{introdc}
F_{DBI}(x)=\sqrt{1-x}
~.
\ee

In what follows we describe how the equations of motion of this system can be reduced to a
first order system for zero-temperature solutions. A general treatment of the finite temperature
case lies outside the immediate scope of this paper. It can be performed along the lines of the
previous subsection. Another interesting generalization suggested by the DBI case would be to
include a $U(1)$ gauge field.

Working again in the domain wall frame we set
\be
\label{introdd}
\dd s^2=e^{2A(u)}(-\dd t^2+\dd x^i \dd x^i)+\dd u^2~, ~~ \phi=\phi(u)
~.
\ee
The equations of motion reduce to the following set of independent differential equations for
the unknown functions $A(u), \phi(u)$
\begin{subequations}
\be
\label{introde}
(d-1) \ddot{A}=V \mathring F {\dot \phi}^2
~,
\ee
\be
\label{introdf}
d(d-1){\dot A}^2-VF+2 V \mathring F {\dot \phi}^2=0
~.
\ee
\end{subequations}
$V$ is a function of $\phi$, $F$ a function of ${\dot \phi}^2$ and we are using the notation
\be
\label{introdg}
\dot{~}:=\frac{\dd}{\dd u}~, ~~
\mathring{F}(x):=\frac{\dd F}{\dd x}
~.
\ee

In this case the superpotential ansatz \eqref{introal}, \eqref{introam} is generalized to
\be
\label{introdi}
\dot \phi=\widetilde W(\phi)~, ~~
\dot A=-\frac{W(\phi)}{2(d-1)}
~.
\ee
Inserting this ansatz into the equations of motion \eqref{introde}, \eqref{introdf} we obtain the first
order system
\begin{subequations}
\be
\label{introdj}
\frac{d}{4(d-1)}W^2-\mathring W \widetilde W-FV=0
~,
\ee
\be
\label{introdk}
\mathring W=-2V\mathring{F} \widetilde W
~.
\ee
\end{subequations}

As an illustrating example consider the DBI case \eqref{introdc}. One can solve explicitly
the second equation \eqref{introdk} to obtain
\be
\label{introdl}
\widetilde W=\frac{\epsilon \mathring{W}}{\sqrt{V^2+{\mathring{W}}^2}}~,~~
F_{DBI}=\frac{V}{\sqrt{V^2+{\mathring{W}}^2}}~, ~~ \epsilon=\pm 1
~.
\ee
Substituting into the first equation \eqref{introdj} we find a single non-linear first order equation
for $W$
\be
\label{introdm}
\epsilon {\mathring{W}}^2+V^2=\frac{d}{4(d-1)} W^2 \sqrt{V^2+{\mathring{W}}^2}
~.
\ee
One of the novelties of the non-linear case is the presence of multiple branches of solutions
parametrized by the free parameter $\epsilon=\pm 1$.

The same results are obtained easily with the use of the Hamilton-Jacobi formalism.
For static and homogeneous configurations Hamilton's function is
\be
\label{introdn}
\SS_H=-M_P^{d-1}\int_{u=u_0} \dd^d x \sqrt{-g}\, W(\phi)
=-M_P^{d-1} \beta V_{d-1} e^{dA_0}W(\phi_0)
~.
\ee
Eq.\ \eqref{introdj} is the Hamilton-Jacobi equation and eq.\ \eqref{introdk} is the first order
flow equation for the scalar field $\phi$.

The formula \eqref{introdn} in the Hamilton-Jacobi formalism implies that the (rescaled) free
energy is
\be
\label{effpotca}
\widehat \FF=-e^{dA_0}W(\phi_0)
~,
\ee
a result that can also be verified easily by direct computation.


\end{document}